\documentstyle[11pt,aaspp4]{article}

\lefthead{Boreiko and Betz}
\righthead{$^{\rm 12}C/$^{\rm 13}$C Isotopic Ratio in M42}

\begin{document}
\def\cii{\ion{C}{2}\ }
\def\thcii{$^{\rm 13}$\ion{C}{2}\ }
\def\twcii{$^{\rm 12}$\ion{C}{2}\ }
\def\ftwo{${\rm F = 2 \leftarrow 1}$}
\def\fonez{${\rm F = 1 \leftarrow 0}$}
\def\foneo{${\rm F = 1 \leftarrow 1}$}
\def\twop{\hbox{$^{\rm 2}{\rm P _{\rm \frac{3}{2}} \leftarrow ^{\rm 2}P _{\rm \frac{1}{2}}}$}}
\def\iso{\hbox{$^{\rm 12}$C/$^{\rm 13}$C}}
\def\oi{\ion{O}{1}\ }
\def\kms{\hbox{km\, s$^{\rm -1}$}}
\def\cmtwo{\hbox{cm$^{\rm -2}$}}
\def\cmthree{\hbox{cm$^{\rm -3}$}}
\def\vlsr{V$_{\rm LSR}$\ }
\def\tr{T$_{r}^{*}$}
\def\tex{T$_{\rm ex}$}
\def\threepone{\hbox{$^{3}P _{1}$\, -\,$^{3}P _{2}$}}
\def\tonec{$\theta ^{1} $C}

\vspace{-0.4in}
\hspace*{\fill}to appear in Ap.J. Letters, Aug. 20, 1996
\vspace{0.3in}

\title{The $^{\rm 12}$C/$^{\rm 13}$C Isotopic Ratio in Photodissociated Gas in M42}

\author{R. T. Boreiko and A. L. Betz}
\affil{Center for Astrophysics and Space Astronomy, University of Colorado,
    Boulder, CO  80309}

\begin{abstract}
We have observed the 158 \micron\ \twop\ fine-structure line of
\twcii simultaneously with the \ftwo\ and \fonez\ hyperfine components
of this transition in \thcii in the Orion photodissociation region near \tonec .
The line profiles were fully resolved using a heterodyne spectrometer with 
0.5 \kms\ resolution.  The relative intensities of these lines give a \iso\
isotopic ratio of $R \, = \,$58 ($^{\rm +6}_{\rm -5}$) for the most probable 
\twcii peak optical depth $\tau \, = \,$1.3 .  The constrained range
of $\tau$($^{\rm 12}$\ion{C}{2})
between 1.0 and 1.4 corresponds to a range of \iso\ between 52 and 61.
The most probable value of 58
agrees very well with that obtained from a relationship
between the isotopic ratio and galactocentric distance derived from
CO measurements, but is lower than the specific value of 67$\pm$3 obtained
for Orion from CO data.
An isotopic ratio as low as 43, as previously suggested based on optical
absorption measurements of the local interstellar medium,
is excluded by the \cii data at about the 2$\sigma$ level.
\end{abstract}


\section{Introduction}
The \iso\ isotopic ratio of the interstellar medium is believed to be
an important parameter for tracing the chemical evolution of the galaxy.
$^{\rm 12}$C is a primary product of stellar nucleosynthesis because it
can be formed in first generation, metal-poor stars, while $^{\rm 13}$C
is a secondary product.  Thus, through successive cycles of star formation,
nucleosynthesis, and enrichment of the interstellar medium with processed
material, the \iso\ ratio is expected to decrease with time, eventually
reaching a steady-state value near 4.  The solar system ratio of $\sim$89
is thought to be representative of the interstellar medium approximately 
5 billion years ago, while the present ratio can serve as a useful check on 
models of galactic evolution.

The \iso\ isotopic ratio in the Orion region has previously been
measured using the isotopomers of many different molecules,
yielding values between 24 and 83, often not overlapping at the 1$\sigma$
level. The data prior to 1980 have been summarized by Wannier (1980), who
finds a formal abundance ratio from these data of 60$\pm$8.
Radio spectra of CH$_{\rm 3}$OH, OCS, HC$_{\rm 3}$N, and the $^{\rm 13}$C
variants of these molecules have been observed in Orion by 
Johansson et al.\ (1984), who
deduced an isotopic ratio near 40.  Langer and Penzias (1990) 
found a relationship between the \iso\ ratio and galactocentric distance
from $^{\rm 13}$C$^{\rm 18}$O and $^{\rm 12}$C$^{\rm 18}$O spectra, but
their measured ratio of 67$\pm$3 in Orion is much higher than would be
expected on the basis of this relationship.  As pointed out by all these
authors, estimating the isotopic ratio from relative intensities
of molecular lines is subject to uncertainty because of the possibilities
of chemical fractionation, self-shielding effects from photodissociating
radiation, and line saturation of the more abundant isotope.  At the
lower radio frequencies, contamination by lines from other species is
not inconsequential at the small signal levels entailed in measurements
of rare isotopomers, especially in the chemically rich Orion IRc2 region.
In addition, uncertainties in relative calibration and pointing when 
spectra are not obtained simultaneously can lead to significant systematic
errors.

In this Letter, we describe well-resolved observations of the \twcii
\twop\ fine-structure transition and the \ftwo\ and \fonez\ hyperfine
components of the corresponding \thcii line near \tonec\ in the Orion 
photodissociation region.  Since these data were obtained
simultaneously, there is no systematic uncertainty caused by
pointing differences or absolute calibration of individual spectra.
Furthermore, since all of the carbon within the PDR is in the form
of C$^{+}$, interpretation of the measured line intensity ratio
to derive the isotopic ratio is relatively straightforward.

\section{Instrumentation and Calibration}
The data were obtained using a far infrared heterodyne receiver 
(\cite{instrument}) flown onboard the Kuiper Airborne Observatory (KAO).
The local oscillator (LO) is an optically pumped CH$_{\rm 2}$F$_{\rm 2}$
laser operating at 1891.2743 GHz (\cite{pse80}),
9.3 GHz away from the rest frequency of the \twcii line at 1900.5369 GHz 
(\cite{cbs86}).
The mixer is a GaAs Schottky diode (University of Virginia type 1T11)
in a corner-reflector mount.  Both the mixer and the first intermediate
frequency (IF) amplifier are cooled to 77 K.  The system
noise temperature measured during the observations was 9500 K single
sideband (SSB).
The IF signal is analyzed by a 400-channel acousto-optic spectrometer (AOS),
with a channel resolution of 3.2 MHz (0.5 \kms ) and a usable bandwidth of 
150 \kms\ at the \cii frequency.

The observations were done using sky chopping at 2.2 and 4.4 Hz, 
with an amplitude of 9\farcm 4 in 1990 and 12\farcm 5 in 1991, respectively.
These chop amplitudes are sufficient to ensure little
emission in the reference beam in the NE-SW chop direction
(see the map of \cite{sta93}).  
The reduction in peak intensity from off-source emission is at most
10\% if there is no velocity gradient in the source, and it is
smaller if the emission in the two reference beams occurs at a different
peak LSR velocity from that in the main beam.  The \cii spectra obtained
in 1990 and 1991 had different reference beam locations because of differences
in both chop amplitude and position angle; thus, the excellent agreement
between line profiles provides further evidence that contamination
from off-source emission is not a significant source of uncertainty
for these observations.  Any residual contamination would affect both
isotopes similarly, and therefore to first order, the derived isotopic
ratio should be unaffected if the optical depth in the \twcii line
is not much greater than 1 (see section \ref{optd}).

The telescope beam size was 43\arcsec\ (FWHM) and the pointing accuracy
is estimated to be $\sim$15\arcsec . Absolute intensity calibration
is derived from spectra of the Moon, for which a physical temperature of
394 K and emissivity of 0.98 are adopted (\cite{lin73}).  Conversion from
double sideband (DSB) to SSB intensity was performed using the calculated
transmission of the atmosphere in the two sidebands,
corrected for the different source and lunar elevation angles.
The observed signals are quoted as
\tr , the antenna temperature corrected for all factors except the
coupling of the source to the beam (\cite{ku81}). The net
calibration uncertainty for a source filling the beam is $\leq$15\% .
The velocity scale accuracy, determined from the known line and LO
rest frequencies, is better than $\pm$0.15 \kms\ (1$\sigma$).

\section{Observations}

The \cii line was observed from the KAO flying at an altitude of 12.5 km on
the nights of 1990 Dec 11 and 1991 Oct 30, Nov 1, and Nov 2.  
Data were obtained at several locations near \tonec\ in M42.  
All the observations were within
35\arcsec\ of the position of peak \cii emission seen in the map of Stacey
et al. (1993), which is approximately 20\arcsec\ SW of \tonec .
The variation in peak intensity between spectra is $\leq$10\%, in agreement
with that expected from the \cii map.  There is, however, a pronounced
velocity gradient over the range covered by our observations,
with the \cii line at \tonec\ redshifted by 1.6 \kms\ relative to that
45\arcsec\ west.  No comparable gradient is seen in the north - south
direction, with peak \vlsr changing by less than 0.2 \kms\ over 40\arcsec .
There is also a significant change in line profile between the \tonec\
spectrum and that seen 45\arcsec\ further west, with the former being wider
and having a $\sim$10\% contribution to the total integrated intensity
from a distinct velocity component characterized by \vlsr \,$= \,$0.8 \kms ,
and the latter showing a somewhat asymmetric profile with an extended
red wing.

All the spectra from positions with integrated intensity above the 90\%
of peak contour from the map of Stacey et al. (1993) were averaged, yielding
a net spectrum with an integration time of 162 minutes.  The corresponding
uncertainty per 0.50 \kms\ channel is 0.37 K (1$\sigma $).
The spectrum includes
two of the three \thcii hyperfine components: \ftwo\ and \fonez ,
at velocities of +11 \kms\ and -65 \kms\ relative to the
\twcii velocity (\cite{cbs86}).  These two components contain 80\% of the total
\thcii fine-structure line intensity.  The \foneo\
hyperfine line containing the remaining 20\% was
outside the bandwidth of the present observations.

\begin{figure}
\centering
\leavevmode
\epsfysize=7.2in
\epsfbox{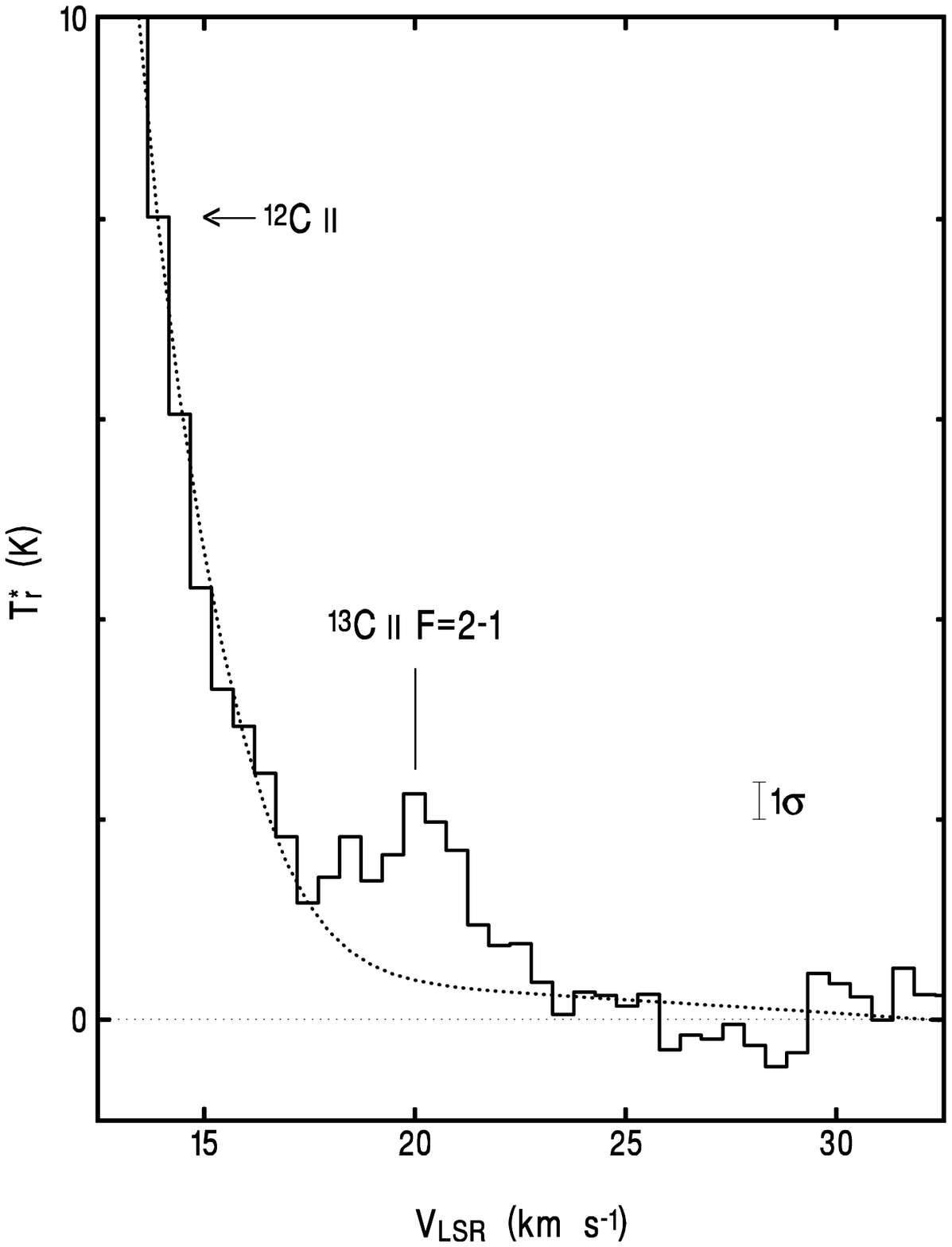}
\caption[fig1.eps]{Observed spectrum of the \ftwo\ hyperfine
component of the 158 \micron\ $^{\rm 13}$C\,{\sc ii} line on the wing of the 
$^{\rm 12}$C\,{\sc ii} line.
The \vlsr\ scale is appropriate only for $^{\rm 12}$C\,{\sc ii}.
The continuum has been removed from the spectrum.
Dotted line is the fit to the $^{\rm 12}$C\,{\sc ii} wing, which was 
subtracted from the data to leave only the $^{\rm 13}$C\,{\sc ii} line.
\label{fit}}
\end{figure}

The \fonez\ \thcii hyperfine line is well-isolated from the \twcii line,
and therefore its intensity can be determined directly.  However, the \ftwo\
component is superposed on the wing of the \twcii line.
Therefore, we fitted the red wing of the \twcii 158~\micron\ line, excluding
the region around the \thcii \ftwo\ frequency, to a sum of Gaussian profiles,
to the statistical uncertainty of the data.  This wing profile was then
subtracted from the spectrum, leaving the desired \thcii\ hyperfine
component.  Figure~\ref{fit} shows an expanded view of the data and the fit
that was subtracted.  No physical significance is attached here to the three
velocity components used in the fit.

\placefigure{fit}

\begin{figure}
\centering
\leavevmode
\epsfysize=7.2in
\epsfbox{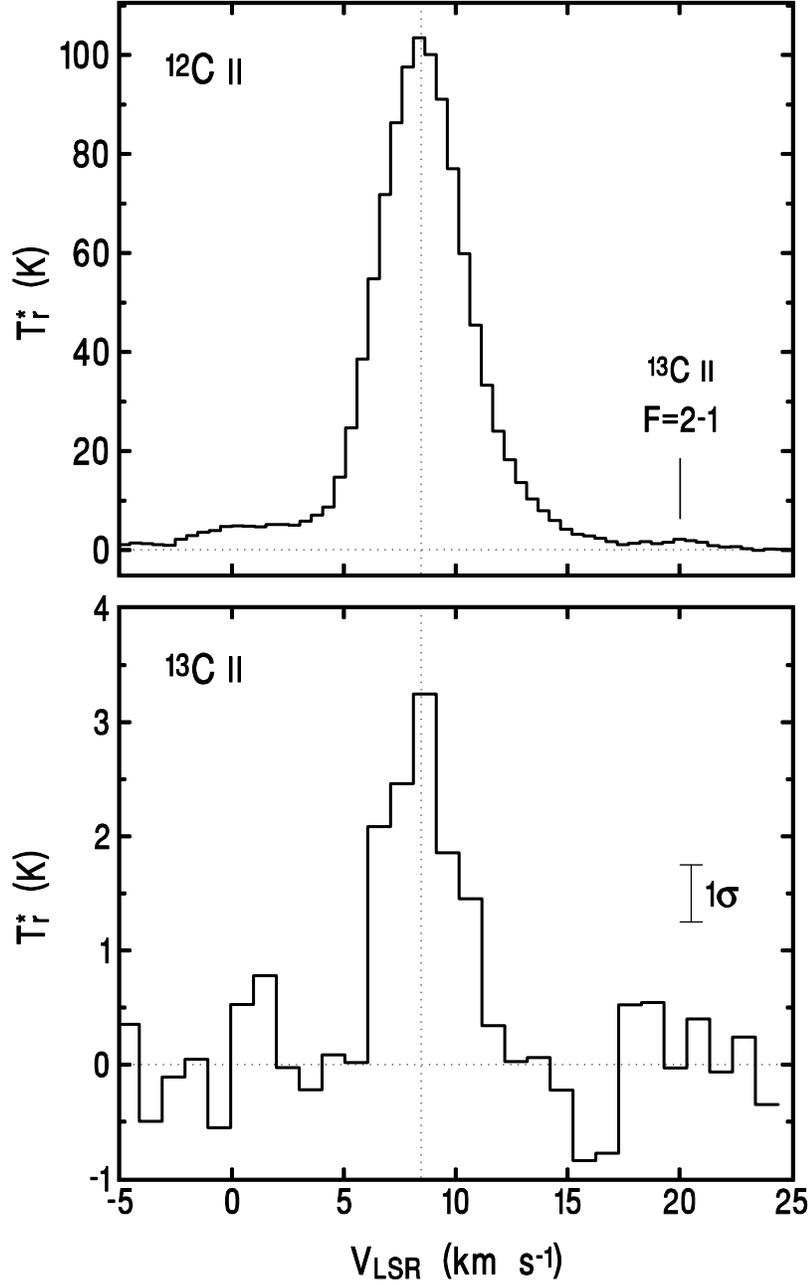}
\caption[fig2.eps]{Observed spectra of the 158 \micron\ 
$^{\rm 12}$C\,{\sc ii}\ and composite $^{\rm 13}$C\,{\sc ii}\ fine-structure 
lines toward \tonec\ (see text for details).  
Integration time is 162 minutes.  For the  $^{\rm 12}$C\,{\sc ii} line,
1$\sigma$ is 0.37 K, too small to be clearly shown in the figure.
The continuum has been removed from both spectra.
\label{12and13}}
\end{figure}

The frequencies for the hyperfine components were refined from the data
by assuming that their velocities agree with that of the \twcii line.
The resulting apparent LSR velocities are 11.6$\, \pm \,$0.2 \kms\ to the red
and 64.5$\, \pm \,$0.4 \kms\ to the blue of the \twcii line for the
\ftwo\ and \fonez\ components respectively, in agreement with those 
given by Cooksy et al. (1986) to within 1$\sigma$.
The spectral regions surrounding the \ftwo\ and \fonez\ \thcii lines were
placed on the correct \vlsr scales and coadded.  The resulting \thcii spectrum
is shown in Figure~\ref{12and13}.  The \tr\ scale has been multiplied by 
1.25 to account for the missing \foneo\ component and therefore represents the 
corrected antenna temperature that would have been measured if all the 
\thcii \twop\ fine-structure radiation had been observed.

\placefigure{12and13}

Table~\ref{tbl1} presents various parameters of the observed spectra.  
The peak \tr , V$_{\rm LSR}$, and width for the \twcii line are derived 
from a Gaussian fit to the line between 5.6 \kms\ and 10.2 \kms\ V$_{\rm LSR}$,
since this narrow part of the line can adequately be described by a Gaussian
profile and appears uncontaminated by other velocity components.
The parameters for the \thcii transitions come from fits to the
entire lines.
For the \thcii \ftwo\ component,  all the parameters are determined
after subtraction of the wing of the \twcii line.
Velocity limits for the integrated
intensity are $\Delta$v(FWHM) on either side of line center for the
\thcii hyperfine components.  However, the limits for the \twcii line
are -5 \kms\ to +25 \kms , in order to include the broad low-intensity emission.
The integrated intensity attributable
to the \thcii \ftwo\ component has been subtracted.  Note that the 
integrated intensity for the \twcii line therefore has a significant
contribution from velocity components other than the main narrow feature
at \vlsr $\sim$8.4 \kms , and there is some broad wing emission beyond the
velocity range used for the calculation.  

\placetable{tbl1}

\section{Analysis and Interpretation}\label{optd}

The ratio of peak intensities for the \twcii and \thcii lines gives the
\iso\ isotopic ratio directly only if both lines are optically thin.  
Models of photodissociation regions suggest that the \twcii line near \tonec\
has a small to moderate optical depth with $\tau \, \lesssim \,$1 
(Tielens and Hollenbach 1985a,1985b).
Stacey et al. (1991) detected the \thcii \fonez\ transition and mapped its
integrated intensity over a small region near \tonec .  These authors calculated
a \twcii line-averaged optical depth $\tau _{av} = $0.75$\pm$0.25 in the inner
region of the nebula (including the immediate vicinity of \tonec )
and $\tau _{av} \, \sim \, $3 for our observed location.  
However, these values depend upon an adopted \iso\ isotopic ratio of 43.

Boreiko, Betz, and Zmuidzinas (1988) observed
the \ftwo\ hyperfine component and derived an optical depth for the
\twcii line near \tonec\ of 5.6$\pm$1.7 using an adopted \iso\ isotopic
ratio of 60.   These earlier data had
a considerably lower signal-to-noise ratio and somewhat lower spectral
resolution than those presented here,
and the \ftwo\ transition was detected only at the 3$\sigma$ level.
While the line position, width, and peak \twcii intensity
observed by Boreiko, Betz, and Zmuidzinas (1988) are in
good agreement with those shown in Table~\ref{tbl1}, the earlier value
for the peak intensity of the \thcii transition is almost a factor of 2
higher, possibly because of inadequate subtraction of the extended wing of
the \twcii line.  The two data sets are in agreement at the 1.5$\sigma$
level, although the difference corresponds to a significant variation
in derived optical depth.

Because gas densities in the \tonec\ region are known to be 
at least ${\rm 10 ^{\rm 5}}$ \cmthree ,
the optical depth of the \twcii line can be determined directly if
the kinetic temperature of the emitting gas is known.  The
63 \micron\ \threepone\ fine-structure line of \oi also arises from
photodissociated gas, and it is optically
thick in models of the Orion PDR (\cite{th85b}).
Boreiko and Betz (1996) fully resolved the \oi and \cii lines at 
\tonec , and they modeled their peak intensities to give
a kinetic temperature
for the photodissociation region of 175\ K - 220\ K, with a
``best-fit'' value of 185\ K.  The uncertainty in the temperature
arises mainly from the unknown density of the region, since the
\oi line may be subthermally excited.  The optical depth of the \twcii line was
found to be between 1.0 and 1.3, with a most likely value of 1.2.

If the kinetic temperature at the observed location is assumed to 
be the same as that at \tonec\ ($\sim$40\arcsec\ further east from the
centroid of observations), then
the optical depth of \twcii is somewhat higher, 1.0 - 1.4, with a most
likely value of 1.3.  This is consistent with the data of Stacey et al. (1991),
which show lower optical depth near the center of the \ion{H}{2} region
surrounding \tonec .
The \iso\ isotopic ratio $R$ can be calculated from the
observed $^{\rm 12}$\ion{C}{2}/$^{\rm 13}$\ion{C}{2} {\it peak} intensity ratio:
\begin{equation}
R = \frac{-\tau _{12}}{{\rm ln } [ 1 - \frac{I_{13}}{I_{12}}
(1 - e ^{-\tau _{12}})]} .
\end{equation}
For a peak optical depth $\tau _{12} \, $=\,1.3 for the \twcii line, 
the relative \twcii and
\thcii peak intensities of Table~\ref{tbl1} give a \iso\ isotopic ratio
of 58\,($^{\rm +6}_{\rm -5}$), where the limits show 1$\sigma$
statistical uncertainty.  The possible range of $\tau _{12}$
between 1.0 and 1.4, determined from the uncertainty in the derived
temperature of the PDR, corresponds to a range of \iso\ between 52 and 61,
with the same statistical uncertainty.

The relative widths of the \twcii and \thcii lines provide supporting
information on the optical depth of the \twcii line.  Since the beamsize
and pointing were identical for all the observed transitions, the
$\sim$20\% difference in FWHM can be ascribed to optical depth broadening
of the $^{\rm 12}$\ion{C}{2} line.  The line profiles are assumed to be 
described by a Gaussian in column density and hence in optical depth.
The intrinsic width is assumed to be the same for the \twcii and
\thcii lines, while their peak optical depths $\tau _{12}$ and $\tau _{13}$
differ by
a factor of $R$, the isotopic ratio.  This factor $R$ does not enter into
the calculation of relative widths from optical depth broadening.
Using line width data only,
$\tau _{12}$\,$=$\,1.1\,($^{\rm +0.9}_{\rm -0.7}$),
in good agreement with the $\tau _{12} \, = \,$1.3 value derived independently
from peak-intensity data.
Spectrally resolved observations of the hyperfine \thcii transitions with
a higher signal-to-noise ratio would be valuable in obtaining a more
precise value of the optical depth of the \twcii line, and, therefore, also
for the \iso\ ratio.

Systematic uncertainties in the kinetic temperature and peak intensity of
the \cii lines affect the derived isotopic ratio in a very nonlinear manner.
Since the \twcii and \thcii components were measured simultaneously, there
is no relative calibration uncertainty.  However, possible self-chopping,
estimated to be below the 10\% level in integrated intensity, will affect
the derived optical depth of the \twcii line.  The kinetic temperature
calculated for the region remains essentially unaffected by self-chopping
of \cii emission.  The main consequence 
would be an increase in the best-fit hydrogen column
density and hence optical depth in the $^{\rm 12}$\ion{C}{2},
$^{\rm 13}$\ion{C}{2}, and \oi lines (see
Boreiko and Betz 1996).  
If the \cii reference beam positions have 10\% of the integrated intensity
of the observed position, then the corrected
peak $^{\rm 12}$\ion{C}{2}/$^{\rm 13}$\ion{C}{2} intensity ratio
would increase to yield an isotopic ratio of 66.

The profile of the \twcii line provides an additional constraint on the
magnitude of any self-chopping.  Since the gas in the off-source beam
is less optically thick than that in the region of interest, self-chopping
(with no velocity gradient) will flatten the peak of the observed line
and increase the apparent width.  Fitting of a Gaussian profile with
moderate optical depth to the \twcii line between 5 and 11 \kms\ 
V$_{\rm LSR}$, where asymmetry and other velocity components are not
significant, suggests that $\tau _{12} \, < \, $1.5 .  At higher
$\tau _{12}$, the reduced $\chi ^{2}$ of the fit exceeds 2 and makes the
data incompatible with an optically thick line profile at the 90\%
confidence level.  Thus, we can conclude that reference beam emission
$\sim$10\arcmin\ from our observed location in the NE-SW chop direction
(which was not sampled by the E-W scan of Stacey et al. 1993)
is either below the 10\% integrated intensity level or occurs at a velocity
at least 1\,-2 \kms\ shifted from that of the peak of the observed line.
High-resolution observations at the reference beam locations would
help to reduce the remaining systematic uncertainty in the optical depth
of the \twcii line and hence in the \iso\ isotopic ratio.

Our value of \iso\ above is in excellent agreement with that obtained from
the relationship of \iso\ with galactocentric radius obtained by Langer and
Penzias (1990) ($R \, = \,$57.4 at the distance of 
Orion), from observations of $^{\rm 12}$C$^{\rm 18}$O
and $^{\rm 13}$C$^{\rm 18}$O.  However, the agreement is not as good
with their value of 79$\pm$7 measured at IRc2 or 67$\pm$3 about 3\arcmin\
NE of IRc2.  As these authors pointed out, the high UV field in the region
could enhance the apparent ratio because of self-shielding of the more abundant 
CO isotope, and therefore these derived values are likely to be too high.
Alternatively, there could be a real difference in \iso\ between the PDR
and the dense molecular cloud cores producing the CO radiation.
We obtained a weak detection of \thcii emission at IRc2, but the
signal to noise ratio of the spectrum is much smaller than at \tonec .
Nevertheless, the derived isotopic ratio of 63\,$\pm$\,22 is
consistent with that at \tonec .

Optical observations of CH$^{+}$ toward nearby bright stars have been
used to obtain a \iso\ isotopic ratio of 43$\pm$4 in the local interstellar
medium with homogeneity at the 12\% level (\cite{hj87}).  However,
molecular observations indicate a higher local value of 60 - 70 (\cite{wm92}),
so either there is significant local inhomogeneity or there remain large
systematic sources of uncertainty in the methods used to determine the
\iso\ ratio.  Our measured value in the PDR of Orion is inconsistent with
the lower value of 43 at approximately the 2$\sigma$ level.

\acknowledgments
We thank the staff of the Kuiper Airborne Observatory for their unfailing
effort and helpfulness during all phases of flight operations.
This work is supported by NASA grants NAG 2-254 and NAG 2-753.

\clearpage
\begin{deluxetable}{lcccc}
\normalsize
\tablecolumns{5}
\tablewidth{0pc}
\tablecaption{{\sc Observed $^{\rm 12}$C\,{\sc ii} and $^{\rm 13}$C\,{\sc ii} Fine-Structure Line Parameters at $\theta ^{1}$C\tablenotemark{a}\label{tbl1}}}
\tablehead{
\colhead{Transition} & \colhead{Peak T$_{r}^{*}$} & \colhead{V$_{\rm LSR}$} & \colhead{Line Width} &
\colhead{Integrated Intensity} \nl 
\colhead{} & \colhead{(K)\ \ \ \ } & \colhead{(km\, s$^{\rm -1}$)}  & \colhead{(km\, s$^{\rm -1}$, FWHM)}  & 
\colhead{(erg\, cm$^{\rm -2}$\, s$^{\rm -1}$\, sr$^{\rm -1}$)}   }
\startdata
$^{\rm 12}$C\,{\sc ii} \twop & 103.43(0.30) & 8.44(0.01) & 4.32(0.03) & 3.80(0.01)$\times$10$^{\rm -3}$\,\tablenotemark{b} \nl
$^{\rm 13}$C\,{\sc ii} \twop\ \ftwo & 1.6(0.2) & 8.9(0.2)\tablenotemark{c} & 3.2(0.5) & 3.98(0.04)$\times$10$^{\rm -5}$ \nl
$^{\rm 13}$C\,{\sc ii} \twop\ \fonez & 0.9(0.2) & 9.1(0.4)\tablenotemark{c} & 4.0(0.9) & 2.75(0.04)$\times$10$^{\rm -5}$ \nl
$^{\rm 13}$C\,{\sc ii} composite\tablenotemark{d} & 3.1(0.3) & 8.4(0.2)\tablenotemark{e} & 3.6(0.4) &  8.35(0.07)$\times$10$^{\rm -5}$ \nl
\enddata
\tablenotetext{a}{Numbers in parentheses represent 1$\sigma$ statistical uncertainties.}
\tablenotetext{b}{Integrated intensity includes several velocity components but not all extended emission.}
\tablenotetext{c}{Calculated using frequencies of Cooksy et al. (1986).}
\tablenotetext{d}{Includes calculated contribution from \foneo\ hyperfine component.}
\tablenotetext{e}{Component frequencies determined from the data, hence \vlsr\ is a defined value.}
\end{deluxetable}

\clearpage


\clearpage

\clearpage

\clearpage

\end{document}